\documentstyle{elsart}
\begin{document}
\begin{pf*}{to be published in: J. Phys. Chem. Sol.}
\begin{frontmatter}
\title{The dimpling in the CuO$_2$ planes \\ 
of YBa$_2$Cu$_3$O$_x$ ($x$\/=6.806-6.984, $T$\/=20-300 K) \\ 
measured by yttrium EXAFS}
\author[Koeln]{J. R{\"o}hler\thanksref{EML}}
\author[Koeln]{S. Link\thanksref{KFA}}
\author[Zuerich]{K. Conder}
\author[Zuerich]{E. Kaldis}
\address[Koeln]{Universit{\"a}t zu K{\"o}ln, 
Z{\"u}lpicherstr. 77, D-50937 K{\"o}ln, Germany}
\address[Zuerich]{Labor f{\"u}r Festk{\"o}rperphysik, ETH-Technopark, 
CH-8005 Z{\"u}rich, Switzerland}
\thanks[EML]{Corresponding author, e-mail: abb12@rs1.rrz.uni-koeln.de}
\thanks[KFA]{Present address: Forschungszentrum J{\"u}lich, IFF, D-52425
J{\"u}lich}
\begin{abstract} 
The dimpling of the CuO$_2$ planes (spacing between the O2,3 and Cu2
layers) in YBa$_2$Cu$_3$O$_x$ has been measured as a function of
oxygen concentration and temperature by yttrium x-ray 
extended-fine-structure spectroscopy (EXAFS). The
relative variations of the dimpling with doping ($x=6.806-6.984$) 
and temperature (20-300 K) are weak (within 0.05 $\rm\AA$), 
and arise mainly from displacements of the Cu2 
atoms off the O2,3 plane towards Ba. The dimpling appears to be
connected with the transition from the underdoped to the overdoped
regimes at $x=6.95$, and with a characteristic temperature in the
normal state, $T^*\simeq 150$ K.

\end{abstract}
%\begin{keyword}
%Superconductivity
%EXAFS
%\end{keyword}
% 
\end{frontmatter}
\section{Introduction}

Coherent structural distortions of the superconducting CuO$_2$ planes
appear to be a crucial issue of the high-$T_c$ cuprates
\cite{PSErice}. The dimpled
basal planes of the CuO$_5$ pyramides in YBa$_2$Cu$_3$O$_x$ emerged 
already from early crystallographic work on this material.
Tilts of the CuO$_6$ octahedra in the La
214 family are frequently considered to be at the origin of
mesoscopic stripe patterns found in some of their compounds. And, as
it is well known, the Bi family of superconductors exhibits long range
order modulations of the CuO$_2$ layers. 
On the other side, the superconducting Hg cuprates 
(exhibiting the highest $T_c$'s) were found to have undistorted 
CuO$_2$ planes. But recent crystallographic work seems 
to converge to relatively weak dimples ($\simeq 0.05$ $\rm\AA$) 
in the basal planes of the CuO$_5$ pyramides of 
the two and more layer Hg compounds \cite{SchKar}. 

The displacements leading to an effective transversal modulation of
the CuO$_2$ planes may be of different kind and physical origin.
''Buckling`` in the La 214 family and the long range order modulations
in the Bi family are understood to arise from tilts of the CuO$_6$
octahedra and the CuO$_5$ pyramides, respectively, both exhibiting
{\em flat} basal planes. On the other hand, the displacements of the
planar Cu and/or planar O atoms along $c$\/ labeled ''dimples`` imply
the basal O--Cu--O bonds to be no longer rigid but soft, and to be
bent in $c$\/ direction. The dimpling in metallic Y-123 is $\geq 0.25$
$\rm\AA$ \cite{ConJil,GuiFur}, but $\leq 0.05$ $\rm\AA$ in metallic
Hg-1212 \cite{SchKar}. We conclude that also other physical properties
than metallicity and superconductivity may be at the origin of the
dimpling, and the net effect possibly attributable to the
superconductivity in the high-$T_c$\/ cuprates may be as small as 0.05
$\rm\AA$. Important consequences from this conclusion arise for both,
experimental and theoretical work.

First, the structural distortions possibly relevant for the
superconductivity in doped Mott insulators depend on the carrier
concentration. Since the doped charge carriers are not expected to be
randomly distributed in the CuO$_2$ planes, the resulting structural
deformations most likely break the translational invariance of the
underlying lattice. Local probes are therefore important experimental
tools for the investigation of this structural problem.

Second, if the dimpling of the CuO$_2$ planes is connected to the
mechanism of superconductivity in the cuprates, it is not justified to
consider their copper oxygen chemistry in the usual way as forming
rigid, strong bonds between the copper atom and the four oxygens.
Consequently theoretical work has to allow for dynamical degrees of
freedom to treat the oxygen and copper orbitals separately. The
theoretical understanding of the experimentally apparent ''soft`` or
bent Cu--O bonds is still at its infancy. But a recent study of the
correlations in a high-$T_c$ cuprate using the Local Ansatz showed
that the charge transfer inside the Cu atoms is much larger than that
between the planar Cu and O atoms, due to correlations \cite{Sto97}.
This finding stresses the need for that atomic correlations of the
copper orbitals ($3d-4s,p$) have to be taken into account to
understand the ''soft`` Cu--O bonding in doped cuprates.

In this brief report we are presenting measurements of the dimpling in
the CuO$_2$ planes of YBa$_2$Cu$_3$O$_x$ by yttrium EXAFS, as a
function of temperature and doping on both sides of the
underdoped-overdoped phase separation line at $x=6.95$.

\section{Experimental Details}
The photoexcited cations in the separating sheet of the two layer
cuprates are ideal observers of the local atomic structure of the
CuO$_2$ planes. We have measured the EXAFS 
beyond the yttrium $K$ edge up to $k\simeq 20$
$\rm\AA^{-1}$ from fine grains ($\leq 5 \mu$m) of YBa$_2$Cu$_3$O$_x$
($x=6.806-6.984$) synthezised under near equilibrium conditons. 
Details of the x-ray spectroscopy, the
data reduction, and of the synthesis of the samples are given
elsewhere \cite{RoeKal}. 

The positions and the vibrational dynamics of the O2,3 and Cu2 atoms
can be extracted from the nearest and next nearest neighbour two-body
scattering configurations Y--O2,3; Y--Cu2, and the nearly collinear
three-body multiple scattering (MS) configurations Y--O2,3--Ba
($\simeq 5$ $\rm\AA$), Y--Cu2--Ba ($\simeq 6.2$ $\rm\AA$). 
To determine the dimpling as a function of temperature we have 
exploited the high sensitivity of these two MS 
configurations to static displacements of the intervening O2,3 and Cu2
atoms, respectively. Since both MS configurations 
include the same Ba-layer, the ratio of their experimental scattering
amplitudes cancels to first order approximation the effects arising
from thermal disorder, and the temperature dependent variations of the
Y--Ba distances. Thus we are able to separate the desired information
on the static positional changes from the thermal disorder, thereby
avoiding weakly based assumptions on the vibrational
dynamics of the Y--Ba cluster.

\section{Experimental Results}
Fig.1 exhibits the temperature and concentration dependencies of the
effective scattering amplitudes of Y--O2,3--Ba (open circles) and
Y--Cu2--Ba (closed circles) for three underdoped ($x=6.806-6.886$) and
two overdoped samples ($x=6.968$, 6.984); $x=6.947$ is close to
optimum doping ($x=6.92$). The average temperature dependencies
exhibit significant deviations from the $1/coth(1/T)$ behaviour
expected from harmonic motions of the diatomic pairs Y--Ba. In
addition dips and step-like features occur in the normal conducting
temperature region. $x=6.886$ exhibits a clear jump at 110 K. We have
carefully checked the sources of systematic errors possibly giving
rise to the discontinuities in the effective scattering amplitudes, in
particular the normalization of the experimental absorption
coefficient. As a result we can definitely exclude experimental
artefacts to be at the origin of the dips in $x=6.806$, 
$x=6.947$, and of the step in $x=6.886$ around 120 K. 
We note that the experimental spectra of $x=6.886$ exhibit the 
relatively best {\em S/N}, while those of $x=6.806$ 
the relatively worst.

The ''raw`` effective scattering amplitudes 
in Fig.1 show directly the dimpling to depend on
both, concentration and temperature. Doping does not significantly
affect the overall behavior of the Y--O2,3--Ba (open circles), 
but moves that of Y--Cu2--Ba (closed
circles, dashed line) for all temperatures clearly beyond Y--O2,3--Ba.
Since the thermally induced damping is almost the same for both
signals, the increasing difference may be straightforwardly attributed
to a relative linearization of Y--Cu2--Ba three body scattering
geometry, {\em i.e.} the Cu2 atoms move off the O2,3 plane towards the
Ba layer.

The temperature dependencies of the dimpling are displayed in Fig.2 in
ascending order from the underdoped to the overdoped regimes. The
numbers given at the left ordinates are from theoretical fits to the 
data using the high order MS approach of the FEFF6 code \cite{ZabEll}. 
The data points, except for x=6.886, exhibit a surprisingly large
scatter rendering unambigous determinations of the detailed 
temperature behaviours difficult. Therefore the superimposed lines
ought to be understood as possible guides to the eyes, among others.
Well known, many copper oxides are notoriously anomalous around
$T\simeq 210$ K and $T\simeq 110$ K \cite{MurRoe}. In fact, the
dimpling exhibits cusps close to these temperatures, however, a
systematic behaviour of these anomalies as a function of doping is
hard to extract. Therefore we ignore them here. The dimpling of
$x=6.806$ and $x=6.886$ may be seen to be independent on temperature
within about $\pm 0.01$ $\rm\AA$. However, the dimpling of underdoped
$x=6.886$ exhibits a shallow but clear minimum around 150 K;
interestingly the step-like feature in Fig.1 at 110 K is almost
cancelled. A minimum around 150 K may be also extracted from nearly
optimum doped $x=6.947$. In the overdoped regime the 150 K minimum
vanishes: the dimpling of $x=6.968$ is almost constant at $T\ge 100$
K, but increases at $T\le 100$ K. Ignoring the strong cusp around 210
K the dimpling of overdoped $x=6.984$ may be seen to increase
linearily from 300 to 20 K.

\section{Concluding Remarks}
The dimples in the CuO$_2$ planes of doped YBa$_2$Cu$_3$O$_x$ and
their variations with oxygen concentration and temperature appear to
be intimately connected with its superconducting properties. The well
known doping-induced increase of $T_c$ is connected with increased
dimples. A significant discontinuity occurs at the
underdoped-overdoped phase separation line ($x=6.95$). Some of us have
recently shown that the discontinuity in the static dimpling
correlates with a sudden drop of the Raman O2,3 in-phase shift
($A_{1g}$ dimpling mode) at $x=6.95$, giving hard evidence for a
new displacive structural phase transformation in the CuO$_2$ planes
\cite{KalLoe}. The temperature dependent data presented in this report
confirm the close connection between the dimpling and the transition
from the underdoped into the overdoped regimes for $20\leq T\leq 300$
K. Moreover, at least for underdoped $x=6.886$, the dimpling appears
to be unambigously connected to a characteristic temperature in the
normal state, $T^{*}\simeq 150$ K.

Interestingly, the variations of the dimpling related with the
dramatic variations of the superconducting properties on doping are
comparatively weak, $\simeq 0.05$ $\rm\AA$, {\em i.e} only $\simeq
20$\% of the total dimpling. Thus its superconductivity related
fraction may be seen to reside on top of an ''offset`` arising from
the particular band structure \cite{AndMaz}.

\ack
Beamtime at the European Synchrotron Radiation Facility (ESRF) 
was under the proposals HC362 and HS45. We thank Dr. P.W. Loeffen for
efficient help setting up appropriately the beamline BL18.

\bigskip
\noindent
{\bf Figure captions}
\bigskip

\noindent
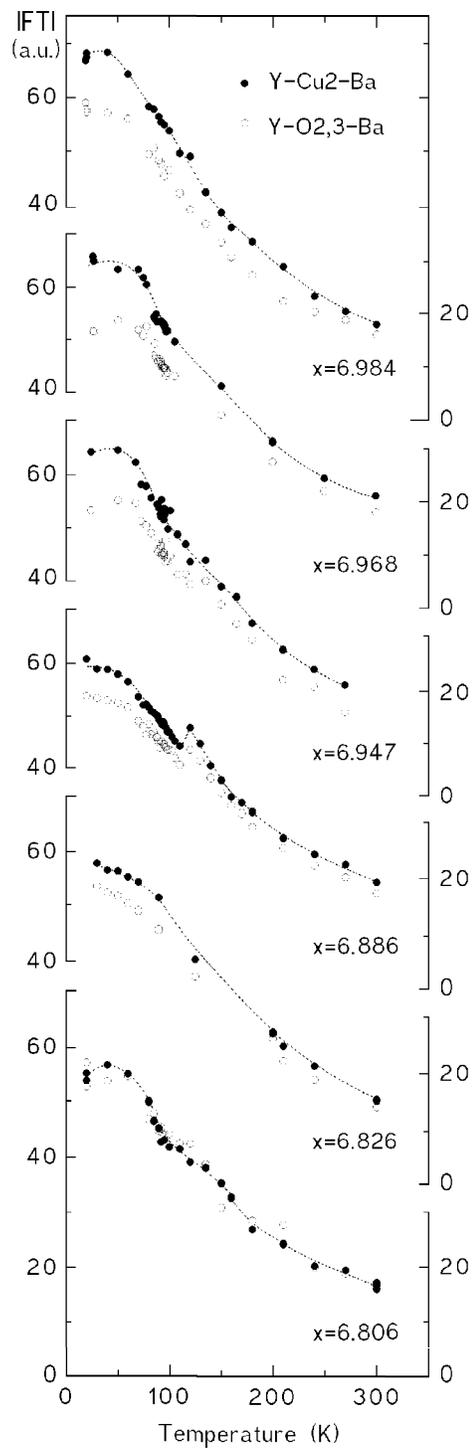
{\bf Fig.1:} Effective scattering amplitudes (arbitrary units) of the
Y--Cu2--Ba (closed circles) and the Y--O2,3--Ba (open circles) MS
con\-fi\-gurations in \\
YBa$_2$Cu$_3$O$_x$ as a function of temperature and
oxygen con\-centration, $x$. \\
Dashed lines are guides to the eyes.
\bigskip

\noindent
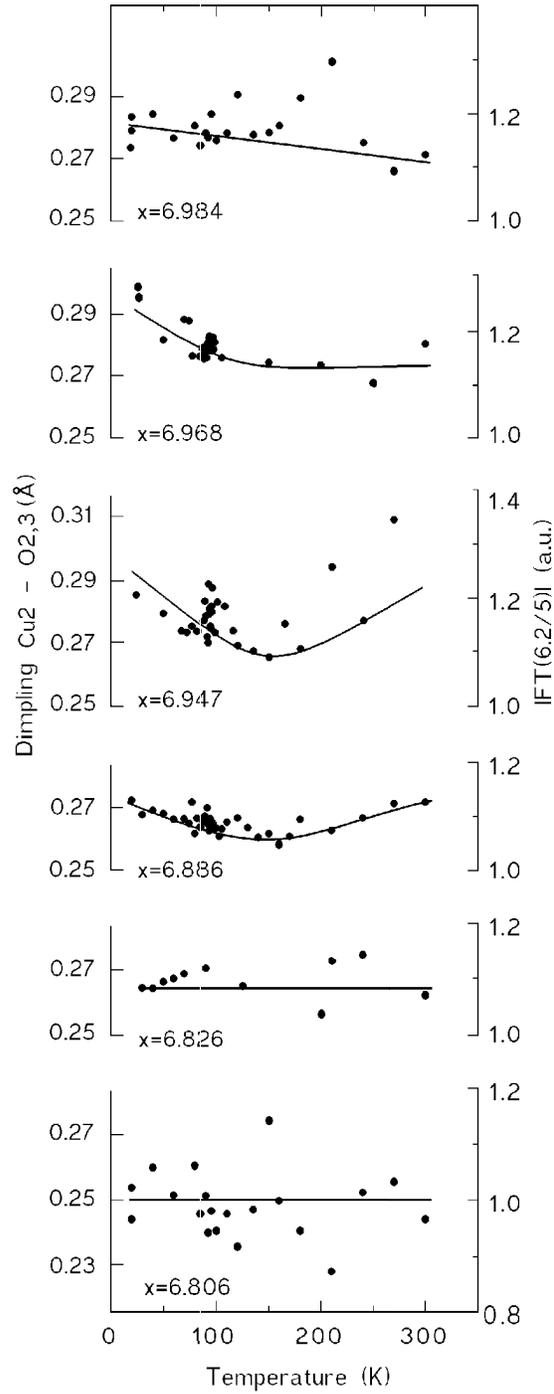
{\bf Fig.2:} Dimpling of the CuO$_2$ planes in YBa$_2$Cu$_3$O$_x$ as a
function of temperature and oxygen concentration, $x$. The superimposed
lines are guides to the eyes. Right hand scale: ratio of the effective
MS amplitudes displayed in Fig. 1.  

\qed
\end{pf*}
\end{document}